# New near-IR observations of mesospheric $CO_2$ and $H_2O$ clouds on Mars

*Mathieu Vincendon[1], Cedric Pilorget[1], Brigitte Gondet[1], Scott Murchie[2], Jean-Pierre Bibring[1]*

[1]Institut d'Astrophysique Spatiale, Université de Paris Sud 11, 91405 Orsay, France (brigitte.gondet@ias.u-psud.fr, phone +33 1 69 85 86 35, fax +33 1 69 85 86 75)

[2]Johns Hopkins University Applied Physics Laboratory, Laurel, Maryland, USA (scott.murchie@jhuapl.edu, phone +1 240-228-6235)

**Abstract:** Carbon dioxide clouds, which are speculated by models on solar and extra-solar planets, have been recently observed near the equator of Mars. The most comprehensive identification of Martian $CO_2$ ice clouds has been obtained by the near-IR imaging spectrometer OMEGA. CRISM, a similar instrument with a higher spatial resolution, cannot detect these clouds with the same method due to its shorter wavelength range. Here we present a new method to detect $CO_2$ clouds using near-IR data based on the comparison of $H_2O$ and $CO_2$ ice spectral properties. The spatial and seasonal distributions of 54 CRISM observations containing $CO_2$ clouds are reported, in addition to 17 new OMEGA observations. CRISM $CO_2$ clouds are characterized by grain size in the 0.5–2µm range and optical depths lower than 0.3. The distributions of $CO_2$ clouds inferred from OMEGA and CRISM are consistent with each other and match at first order the distribution of high altitude (>60km) clouds derived from previous studies. At second order, discrepancies are observed. We report the identification of $H_2O$ clouds extending up to 80 km altitude, which could explain part of these discrepancies: both $CO_2$ and $H_2O$ clouds can exist at high, mesospheric altitudes. CRISM observations of afternoon $CO_2$ clouds display morphologies resembling terrestrial cirrus, which generalizes a previous result to the whole equatorial clouds season. Finally, we show that morning OMEGA observations have been previously misinterpreted as evidence for cumuliform, and hence potentially convective, $CO_2$ clouds.

# 1. Introduction

Carbon dioxide ice clouds form on Mars at high latitudes during the polar night (Ivanov & Muhleman 2001; Neumann, Smith & Zuber 2003; Forget, Hansen & Pollack 1995; Giuranna et al. 2008; Pettengill & Ford 2000) and at equatorial latitudes during daytime (Montmessin et al. 2007; Määttänen et al. 2010) and probably nighttime (Clancy & Sandor 1998; Schofield et al. 1997; Montmessin et al. 2006). $CO_2$ clouds may have contributed to warm the surface of early Mars (Forget & Pierrehumbert 1997) and have to be taken into account while estimating the climate – an hence the habitability – of extrasolar planets (Forget & Pierrehumbert 1997; Selsis et al. 2007). The formation of $CO_2$ clouds has also been considered on Earth within the framework of global glaciations (Caldeira & Kasting 1992). Assessing the actual impact of $CO_2$ clouds on the climate require precise constraints about their physical properties and formation mechanisms (Colaprete & Toon 2003; Glandorf et al. 2002; Forget & Pierrehumbert 1997; Mischna et al. 2000), and Mars is currently the only known terrestrial planet where $CO_2$ clouds can be studied.

Studies presenting indirect pieces of evidence for $CO_2$ clouds have been numerous over the past decades and have been at the origin of the idea that $CO_2$ clouds can form both at polar and equatorial latitudes (Bell et al. 1996; Clancy et al. 2007; Montmessin et al. 2006; Clancy & Sandor 1998; Schofield et al. 1997; Clancy et al. 2004). These studies were mainly based on the observation of clouds at altitude where temperature is cold enough for $CO_2$ to condense (Montmessin et al. 2006; Clancy & Sandor 1998; Clancy et al. 2007) and/or where the amount of water vapor is expected to be inconsistent with the observed cloud thicknesses (McConnochie et al. 2010; Clancy et al. 2007). The ambiguity of these tentative identifications has several origins, such as possible confusion with $H_2O$ clouds (Clancy et al. 2007) or surface ice (Bell et al. 1996). Mars Express has provided the definitive spectroscopic evidence for the existence of $CO_2$ ice particles in the atmosphere of Mars (Formisano et al. 2006; Montmessin et al. 2007). In particular, the near-IR imaging spectrometer OMEGA has provided the first spectroscopic identification of a cloud as being composed of $CO_2$ ice.

Recent studies have provided significant information about the physical properties and distribution of $CO_2$ clouds. Martian equatorial $CO_2$ clouds form at mesospheric altitudes of about 50 to 100 km (Määttänen et al. 2010; Scholten et al. 2010; Montmessin et al. 2007; Montmessin et al. 2006; Clancy et al. 2007). They are characterized by visible optical depths ≲ 0.5 and mean grain size ("effective radius") ≲ 3 µm (Montmessin et al. 2007; Määttänen et al. 2010; McConnochie et al. 2010; Clancy et al. 2007). These equatorial clouds are observed at specific time and location, mostly between 20°S and 20°N near Valles Marineris and Terra Meridiani for solar longitudes in the [0° - 140°] range (Määttänen et al. 2010; Clancy et al. 2007). On the contrary, high latitudes clouds found during the polar night form in the lower 20 km of the atmosphere and are probably characterized by larger optical depths and grain sizes (Forget, Hansen & Pollack 1995; Glandorf et al. 2002; Ivanov & Muhleman 2001; Colaprete, Haberle & Toon 2003). A few occurrences of mid-latitudes (40° – 50°) clouds have also been reported in both the northern and the southern hemisphere during local autumn (Montmessin et al. 2007; Määttänen et al. 2010; Scholten et al. 2010; McConnochie et al. 2010).

Low to mid-latitudes clouds can be more easily observed and studied than polar clouds as they occur during daytime. Despite the large number of $CO_2$ clouds observations acquired so far, several properties are still subject to debate as they differ depending on the dataset, method or viewing geometry used for their estimation. For example, the four nighttime clouds observed in SPICAM limb data and expected to be $CO_2$ clouds (Montmessin et al. 2006) are thin ($\tau < 0.01$), small-grained ($r_{eff} \simeq 0.1$ µm) and located for half of them in the 30°S – 40°S latitude range, which is not consistent with daytime nadir direct detections of $CO_2$ clouds by OMEGA (Määttänen et al. 2010). The discrepancy could be due to day-night variability (González-Galindo et al. 2009) or to observational/methodological biases (nadir versus limb viewing geometry / confusion with $H_2O$ clouds). Similarly, clouds interpreted as probable $CO_2$ clouds have been reported in the Ls 140° - 170° range from TES/MOC limb data (Clancy et al. 2007) while $CO_2$ cloud detections are almost never reported at these $L_S$ from OMEGA nadir data (Määttänen et al. 2010). The cloud morphology is also subject to debate as it has been alternatively described as mostly "cumuliform" (Montmessin et al. 2007) or on the contrary mainly "ripple-like to filamentary" (Scholten et al. 2010). This apparent disagreement could be due to observational biases such as spatial sampling or to the existence of different categories of clouds possibly related to local time or solar longitude (Määttänen et al. 2010). The cumuliform morphology of $CO_2$ clouds inferred from some observations has raised the question of a possible convective activity within these clouds (Montmessin et al. 2007; Määttänen et al. 2010; Colaprete et al. 2008). Calculation have however concluded that the whole energy available in the cloud – in the form of latent heat – can form at maximum a convective cell 3 to 5 order of magnitude smaller than the size of the cloud (Määttänen et al. 2010). Getting better constraints about the distribution and properties of $CO_2$ clouds will help to understand their formation mechanism: on the one hand, ongoing studies try to assess which physical mechanisms can decrease mesospheric temperatures (Spiga et al. 2010; González-Galindo et al. 2010) as global circulation models do not indicate sufficiently cold temperatures to initiate $CO_2$ cloud (Montmessin et al. 2007); one the other hand, a better understanding of the microphysics of $CO_2$ cloud formation is required (Määttänen et al. 2010), notably to understand why subfreezing temperatures are observed without clouds (Forget et al. 2009).

OMEGA (Bibring et al. 2005) is a very powerful tool for the study of $CO_2$ clouds as it is able to unambiguously identify the $CO_2$ composition of a cloud based on a near-IR spectral feature located at 4.26 µm (Montmessin et al. 2007; Määttänen et al. 2010). OMEGA is however not exempt of observational biases, such as a low spatial resolution of 0.3 to 5 km and gaps in the spatial/time coverage. Moreover, the "4.26 µm" method cannot detect potential low altitude or small-grained $CO_2$ clouds (Montmessin et al. 2007). The CRISM instrument is – with OMEGA – the second visible and near-IR imaging spectrometer currently orbiting Mars. CRISM offer a significantly higher spatial resolution (20 meters) than OMEGA but its shorter wavelength range does not include the diagnostic "$CO_2$ ice" 4.26 µm feature.

In this paper, we present a new method that makes it possible to identify $CO_2$ clouds in the CRISM dataset, without ambiguity regarding $H_2O$ clouds. Results are compared to those established by previous studies, which make it possible to extend our knowledge of $CO_2$ cloud occurrences and properties, and understand the different observational biases associated with various data sets. New nadir and limb observations by OMEGA are also presented.

## 2. CRISM data analysis

### a. Identification of clouds

The CRISM web map interface (http://crism-map.jhuapl.edu/) provides a rapid overview of the spectral images acquired by CRISM. For each targeted observation, a visible RGB composite image based on wavelengths 0.592 µm, 0.533 µm and 0.492 µm and a near-IR brightness image at a wavelength of 1.33 µm are automatically processed (Murchie et al. 2007). We have performed a global survey of all CRISM targeted observations obtained from the beginning of the mission to summer 2010 in the latitude range [60°S – 60°N] using this interface. Both the visible and near-IR images have been scanned in search of cloud morphologies. About 200 observations have been found to show morphological features characteristic of atmospheric clouds (see examples in Figure 1). Cloudy features appear systematically in the RGB images but are frequently poorly discernable in the near-IR. Observational biases associated with this detection method are discussed in section 2.b and 2.d.

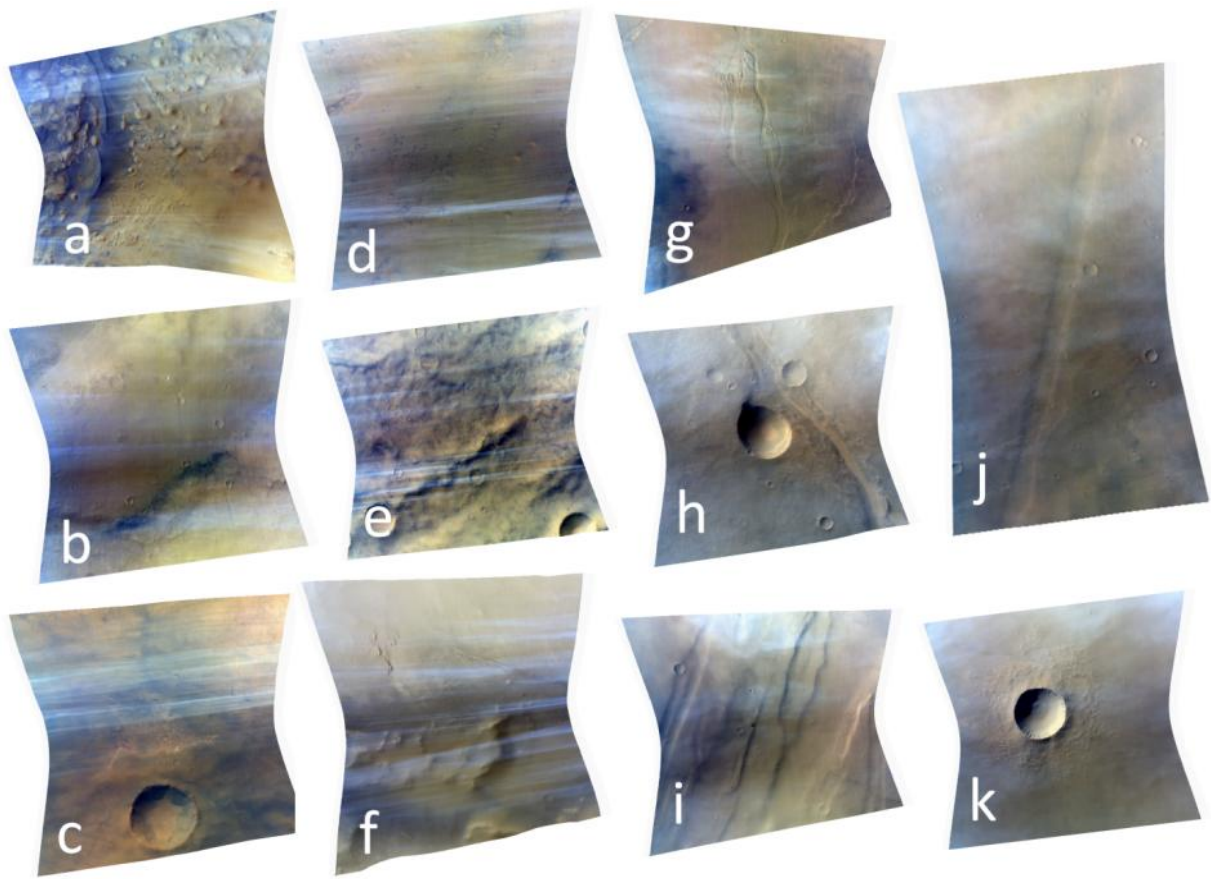

*Figure 1: Examples of CRISM (Murchie et al. 2007) visible images (RGB composite) with clouds. North is on top. Details about these observations are provided in Table 1 and Table 2. From spectroscopy, images a to f correspond to $CO_2$ clouds while images g to k correspond to $H_2O$ clouds (see text for details). At the surface, images are about 11x11 km wide. A given CRISM "targeted" image is acquired with a varying emergence angle between ±35° from south to north (Murchie et al. 2007). As a consequence, important geometric distortions occur while looking at features that are not located at ground altitudes. For example, clouds located at an altitude of 75 km will be compressed by 10 in the north-south direction without change in the east-west direction (see Figure 5 for details).*

*Table 1: list and properties of CRISM observations containing $CO_2$ clouds. Observations used as examples in figures 1, 3, 13, 14, 15 and 16 are indicated in the last column. "app. morph." refers to the apparent morphology classification performed in Figure 14.*

| observation | year_doy | latitude (°) | longitude (°E) | Ls (°) | app. morph. | Figure |
|---|---|---|---|---|---|---|
| FRT00009509 | 2008_010 | -7.57 | -96.11 | 15.370 | 1 | |
| FRT000096EE | 2008_015 | -10.44 | -98.45 | 17.812 | 2 | |
| FRT00009905 | 2008_021 | -7.37 | -96.02 | 20.715 | 1 | Figure 1a |
| FRT00009A6E | 2008_026 | -12.87 | -97.16 | 23.122 | 3 | |
| FRT00009BB | 2008_031 | -7.00 | -99.09 | 25.516 | 1 | |
| FRT0000991A | 2008_021 | -5.78 | 154.64 | 20.862 | 1 | |
| FRT000097F0 | 2008_018 | -7.28 | 25.78 | 19.101 | 1 | |
| FRT0000A344 | 2008_062 | -3.65 | 23.65 | 39.427 | 1 | Figure 1f |
| FRT0000A125 | 2008_054 | -9.65 | 11.19 | 35.731 | 1 | |
| FRT0000A3D7 | 2008_064 | -8.05 | 6.55 | 40.372 | 1 | Figure 1e, 14 |
| FRT0000931E | 2008_004 | 0.99 | 14.25 | 12.761 | 2 | |
| FRT000095B0 | 2008_011 | 8.61 | 21.07 | 16.187 | 2 | |
| FRT00009A59 | 2008_026 | -5.45 | 11.46 | 22.978 | 1 | |
| FRT00009BD4 | 2008_032 | 4.00 | 13.35 | 25.842 | 1 | Figure 1b |
| FRT00009E67 | 2008_043 | 4.00 | 13.35 | 31.047 | 1 | Figure 15, 3 |
| HRL000094EF | 2008_009 | -3.75 | 12.19 | 15.221 | 2 | |
| HRL00009731 | 2008_016 | 8.85 | 13.67 | 18.146 | 1 | |
| FRT000092C5 | 2008_003 | 1.71 | 8.84 | 12.274 | 1 | |
| FRT0000966C | 2008_014 | -0.12 | 4.39 | 17.186 | 1 | |
| FRT0000984F | 2008_019 | -1.84 | 3.18 | 19.616 | 1 | Figure 1c |
| FRT00009A0F | 2008_025 | 3.18 | 4.58 | 22.505 | 3 | |
| FRT00009B5A | 2008_030 | 4.46 | 2.81 | 24.902 | 2 | |
| HRL0000AC53 | 2008_131 | 2.41 | 3.71 | 69.820 | 3 | |
| FRT0000942E | 2008_007 | 0.33 | 1.62 | 14.254 | 1 | |
| FRT00009C73 | 2008_035 | 3.51 | 0.70 | 27.284 | 1 | |
| HRL00008C6B | 2007_350 | 3.74 | -0.63 | 3.272 | 2 | |
| FRT00009B1C | 2008_029 | 2.13 | -1.23 | 24.431 | 1 | |
| FRT00016F45 | 2010_061 | 2.23 | -1.31 | 58.778 | 2 | |
| FRT000097F8 | 2008_018 | 1.83 | -2.36 | 19.139 | 1 | |
| FRT00009985 | 2008_023 | 0.00 | -4.46 | 21.557 | 1 | Figure 1d |
| FRT000095B8 | 2008_011 | -2.05 | -5.48 | 16.223 | 1 | |
| FRT0000979C | 2008_017 | -3.25 | -7.72 | 18.660 | 1 | |
| FRT0000BD84 | 2008_212 | -3.45 | -7.81 | 105.679 | 1 | Figure 13 |
| FRT0000C71F | 2008_256 | -2.99 | -7.93 | 126.146 | 1 | |
| FRT00009554 | 2008_010 | -2.96 | -9.84 | 15.741 | 1 | |
| HRL000095C7 | 2008_012 | 1.11 | -59.84 | 16.298 | 2 | |
| FRT0000CBC3 | 2008_273 | -4.23 | -62.82 | 134.133 | 1 | |
| FRT00009901 | 2008_021 | -6.28 | -69.13 | 20.679 | 1 | |
| FRT00009A67 | 2008_026 | -6.38 | -70.48 | 23.087 | 1 | Figure 16 |
| FRT00009CE4 | 2008_037 | -6.26 | -69.14 | 28.326 | 1 | |
| FRT0000CCB5 | 2008_277 | -6.26 | -69.36 | 136.150 | 3 | |
| HRL0000CD5A | 2008_280 | -10.02 | -53.78 | 137.642 | 3 | Figure 14 |
| FRT0000C67C | 2008_253 | -9.90 | -77.29 | 124.777 | 1 | |
| FRT000171D9 | 2010_069 | -12.10 | -73.27 | 62.019 | 1 | |
| HRS0000A076 | 2008_051 | -12.51 | -85.42 | 34.457 | 1 | |
| FRT00009998 | 2008_023 | -8.38 | -84.92 | 21.665 | 2 | |
| FRT00009C50 | 2008_034 | -7.22 | -84.46 | 26.923 | 1 | |
| FRT0000A0BC | 2008_052 | -8.00 | -80.97 | 34.918 | 1 | |
| FRT00009DF9 | 2008_041 | -5.81 | -76.00 | 30.224 | 2 | |
| FRT0000BB2A | 2008_204 | -5.35 | -75.32 | 102.060 | 1 | |
| FRT00009CB6 | 2008_036 | -5.83 | -73.87 | 27.859 | 2 | Figure 14 |
| HRL0000C59C | 2008_249 | -3.78 | -71.71 | 122.825 | 1 | |
| FRT0000999D | 2008_023 | -8.37 | -112.77 | 21.702 | 3 | |
| FRT0000CD9F | 2008_281 | -1.91 | -103.68 | 138.219 | 1 | |

*Table 2: list and properties of CRISM observations containing $H_2O$ clouds.*

| observation | year_doy | latitude (°) | longitude (°E) | Ls (°) | app. morph. | Figure |
|---|---|---|---|---|---|---|
| FRT0000C428 | 2008_243 | 18.80 | 5.22 | 119.828 | 5 | |
| FRT0000C8C8 | 2008_261 | 21.98 | 14.64 | 128.566 | 5 | |
| HRL0000A316 | 2008_061 | 3.94 | -91.93 | 39.112 | 5 | Figure 1j |
| FRT0000AA85 | 2008_121 | 1.73 | -73.27 | 65.429 | 5 | |
| FRT0000A28C | 2008_059 | -11.30 | -99.46 | 38.197 | 5 | |
| FRT0000B7DA | 2008_194 | -7.14 | -98.92 | 97.492 | 4 | |
| HRL0000A438 | 2008_065 | -4.50 | -98.14 | 40.967 | 5 | |
| HRL0000B150 | 2008_166 | -13.22 | -97.88 | 85.223 | 5 | |
| HRL0000BF61 | 2008_221 | -6.97 | -98.95 | 110.000 | 5 | |
| HRL0000A452 | 2008_065 | 21.53 | 149.39 | 41.110 | 5 | |
| HRL0000B660 | 2008_189 | 19.36 | 150.32 | 95.344 | 5 | |
| FRT000043DB | 2007_042 | -13.18 | 93.19 | 182.264 | 4 | |
| HRL0000C2A0 | 2008_236 | 15.94 | 56.57 | 116.888 | 5 | |
| HRL0000AE10 | 2008_156 | 22.37 | 43.25 | 80.989 | 5 | |
| HRS0000AC0B | 2008_129 | 15.01 | 52.51 | 69.311 | 5 | |
| HRS0000B3DA | 2008_178 | 18.88 | 39.84 | 90.477 | 5 | Figure 3, 14 |
| HRL00013EC0 | 2009_205 | -16.91 | 52.09 | 309.256 | 4 | |
| HRL0000CCF1 | 2008_278 | -1.91 | -9.62 | 136.571 | 4 | Figure 14 |
| FRT0000A1B4 | 2008_056 | -0.00 | -61.91 | 36.755 | 4 | |
| FRT00009562 | 2008_011 | -5.03 | -63.86 | 15.815 | 5 | |
| HRL0000CE4F | 2008_285 | -20.67 | -53.97 | 140.182 | 5 | |
| FRT0000B2D9 | 2008_173 | -13.07 | -90.16 | 88.377 | 5 | Figure 1k |
| HRL0000C228 | 2008_235 | -12.51 | -86.20 | 116.126 | 5 | |
| HRS0000BA54 | 2008_201 | -11.83 | -89.76 | 100.696 | 5 | |
| FRT0000AF87 | 2008_161 | -15.93 | -94.98 | 82.966 | 5 | Figure 1h |
| FRT0001706A | 2010_065 | 13.28 | -97.12 | 60.247 | 5 | |
| HRL0000AB99 | 2008_127 | 19.30 | -99.80 | 68.154 | 5 | |
| HRS00003756 | 2006_353 | 15.01 | -98.06 | 153.041 | 5 | |
| FRT00009B6E | 2008_030 | 17.56 | -107.81 | 25.048 | 5 | |
| FRT0000A0F8 | 2008_053 | 6.17 | -105.64 | 35.414 | 4 | |
| FRT0000A56D | 2008_069 | 16.52 | -107.97 | 42.820 | 5 | |
| FRT00016FB1 | 2010_063 | 15.87 | -107.42 | 59.360 | 5 | |
| FRT0000BF03 | 2008_219 | 17.43 | -111.22 | 109.078 | 5 | |
| FRT0001685C | 2010_045 | 21.90 | -111.21 | 51.684 | 5 | Figure 1g |
| FRT00016F5E | 2010_062 | 20.17 | -112.46 | 58.916 | 5 | |
| FRT000172A4 | 2010_071 | 6.92 | -120.63 | 62.975 | 5 | |
| HRS0000A91E | 2008_114 | 20.16 | -110.78 | 62.331 | 5 | |
| FRT00009CC0 | 2008_037 | 17.84 | -133.09 | 27.933 | 4 | |
| FRT0000D333 | 2008_307 | -2.02 | -113.41 | 151.163 | 4 | |
| FRT000137B8 | 2009_185 | -8.75 | -105.31 | 297.330 | 3 | |
| FRT00014210 | 2009_217 | -3.89 | -113.23 | 316.071 | 5 | |
| FRT0001718C | 2010_068 | -8.76 | -105.32 | 61.610 | 5 | |
| HRS00003F18 | 2007_018 | 0.39 | -112.57 | 168.878 | 5 | |
| FRT0000408B | 2007_027 | -9.16 | -120.23 | 173.967 | 5 | |
| FRT0000A4DC | 2008_067 | -2.39 | -115.40 | 41.911 | 5 | |
| FRT0000A07D | 2008_051 | 4.79 | -114.63 | 34.494 | 4 | |
| FRT0000912A | 2007_363 | -9.00 | -124.64 | 9.979 | 4 | |
| FRT00013C47 | 2009_198 | -9.49 | -125.72 | 304.721 | 4 | |
| HRL0000AAB6 | 2008_122 | 5.66 | -122.97 | 65.940 | 5 | |
| FRT0000B421 | 2008_179 | -17.55 | -113.51 | 91.127 | 5 | |
| FRT0000ACFD | 2008_153 | -14.43 | -107.87 | 79.383 | 5 | |
| FRT00008854 | 2007_303 | -10.00 | -121.98 | 339.176 | 3 | |
| FRT0000933E | 2008_005 | -9.90 | -121.13 | 12.947 | 4 | |
| FRT00017AC1 | 2010_093 | 2.54 | 46.54 | 72.640 | 4 | |
| FRT00018D05 | 2010_136 | 2.13 | -100.30 | 91.295 | 5 | |
| FRT000180E9 | 2010_108 | -7.16 | -98.90 | 79.106 | 4 | |
| FRT000173B5 | 2010 _074 | -4.55 | -103.07 | 64.303 | 5 | Figure 1i |

| | | | | | |
|---|---|---|---|---|---|
| FRT00018745 | 2010_121 | 7.31 | -118.01 | 84.980 | 4 |
| FRT00018C5F | 2010_134 | 9.08 | -111.49 | 90.401 | 5 |
| HRL00018570 | 2010_117 | 8.89 | -111.08 | 83.168 | 5 |
| HRL00018A37 | 2010_127 | 8.62 | -113.92 | 87.687 | 5 |
| FRT00017245 | 2010_070 | -6.75 | -123.42 | 62.531 | 5 |
| FRT0001734E | 2010_073 | -14.92 | -107.48 | 63.859 | 5 |
| FRT00016864 | 2010_045 | 16.92 | -138.91 | 51,718 | 5 |
| FRT00017468 | 2010_076 | 15.97 | -96.58 | 65.190 | 5 |
| FRT000176AE | 2010_082 | 18,16 | -93.66 | 67,879 | 5 |
| FRT00017C60 | 2010_098 | 13.27 | -99.03 | 74.614 | 5 |
| HRS000183D9 | 2010_114 | 21.52 | -72.89 | 81.769 | 5 |
| FRT00018A18 | 2010_127 | 19.51 | -33.94 | 87.585 | 5 |
| FRT0001676B | 2010_043 | 23.72 | 68.62 | 50.548 | 5 |
| HRL000166FE | 2010_041 | 24.14 | 144.18 | 49.997 | 5 |
| HRL00018666 | 2010_119 | 23.54 | 148.80 | 84.195 | 5 |

### b. Cloud spectral properties

We first analyze the spectral properties of these clouds to estimate their composition. Observations obtained above a cloud contain both the cloud and the surface signatures (the latter frequently dominates). To isolate the cloud impact on the radiance, we need to compare the spectral properties observed above a cloud to the spectral properties of a nearby surface assumed to be similar to the surface below the cloud. CRISM clouds are typically divided in small optically thicker areas which extent is lower than a few km for the north-south dimension (Figure 1). CRISM observations are obtained at about 15.00 local time, hence the sun is typically SW oriented with a solar zenith angle of about 45° near the equator. Considering that the altitude of clouds is greater than a few km, the surface below a given area of the cloud is mainly illuminated by sunlight that did not go through it. At first order, the reflectance $R_{Cloud}$ measured by CRISM above a cloud is thus of the form (see Figure 2):

$$R_{Cloud} = R_{Surface} \times T_{Cloud} + S_{Cloud}$$

where $R_{Surface}$ is the reflectance measured above the nearby surface (assumed to be Lambertian), $T_{Cloud}$ is the cloud transmission factor (affecting the light coming from the ground and going through the cloud), and $S_{Cloud}$ is the reflectance of the cloud (light scattered upward by the cloud without interaction with the surface). Note that $T_{Cloud}$, which account for multiple scattering within cloud particles, can be greater than 1. This simple equation neglects the role of the light scattered downward by the cloud and then scattered upward by the surface through the cloud, a relevant approximation considering the small extent of the clouds compared to their altitudes. To first order, the ratio $R_{Cloud}/R_{Surface}$ is an estimate of $T_{Cloud}$ when $R_{Surface}$ is sufficiently high, and the subtraction $R_{Cloud} - R_{Surface}$ is an estimate of $S_{Cloud}$ when $T_{Cloud}$ is close to 1 and/or $R_{Surface}$ is close to 0. As observations provide both $R_{Cloud}$ and $R_{Surface}$, we can therefore rapidly infer an estimate of $T_{Cloud}$ and $S_{Cloud}$ by calculating the ratio and subtraction of the cloud spectrum versus the surface spectrum (see an example in Figure 3). These estimates will be compared to radiative transfer modeling of $T_{Cloud}$ and $S_{Cloud}$ in section c.

We assume that only 3 components can mainly participate to the composition of present day Martian clouds given our current comprehension of the physics and chemistry of the Martian atmosphere: mineralogical aerosols ("Martian dust"), water ice, and $CO_2$ ice. Other species, such as H$_2$SO$_4$, have been hypothesized to condense in the Martian atmosphere during earlier epochs (Levine & Summers 2011). However, there is no observational or theoretical evidence for the formation of these aerosols on present-day Mars. Metastable particles of clathrate could theoretically exist (Chassefière 2009), but have not been observed and are not expected to form localized clouds as they would originate from the subsurface where the pressure is high enough for their formation. Spectral signatures of such clathrate hydrate particles (methane clathrate, CO$_2$ clathrate) will be dominated by their main component: water ice (Schmitt, Mulato & Douté 2003; Dartois & Schmitt 2009). Micrometeorites, which spectral properties will differ from Martian dust, are clearly not expected to form localized clouds of small particle size such as observed by CRISM (Court & Sephton 2011; Toppani et al. 2001).

Martian dust aerosols can be easily distinguished from ice particles in the visible wavelengths: both CO$_2$ and H$_2$O ices present no major absorptions ($T_{Cloud}$ is close to unity) while Martian dust is characterized by a strong, ubiquitous ferric absorption with $T_{Cloud}$ decreasing from 1 µm to 0.4 µm (Vincendon et al. 2009). We have detected no significant contribution from dust for the clouds analyzed in this study. This is consistent with the method used to identify clouds (see section 2.a): ice is bright at visible wavelengths compared to the surface, while the spectral differences between surface and dust aerosols are smaller as they both contain a ferric absorption feature. We present in the next section a methodology to differentiate CO$_2$ from H$_2$O clouds on the basis of their apparent spectral properties.

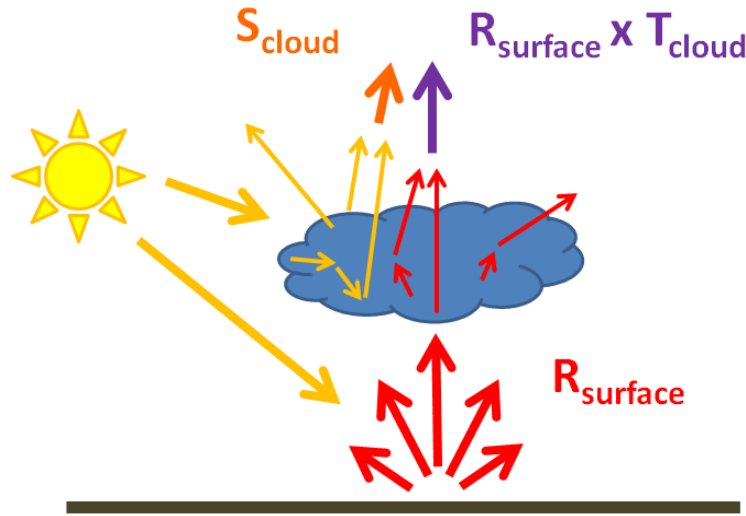

*Figure 2: Radiative transfer scheme considered. The ground below the cloud is illuminated by "direct" sunlight (no attenuation by the cloud itself) as we consider small clouds (see text). The reflectance $R_{Cloud}$ measured above the cloud is thus composed of the scattered reflectance of the cloud $S_{Cloud}$ (orange arrow, composed of single and multiple scattering by cloud particles without interaction with the surface), plus the reflectance coming from the surface (red arrow) modulated by the cloud transmission factor which account for multiple scattering between cloud particles: $R_{Surface} \times T_{Cloud}$ (purple arrow).*

### c. Spectral discrimination between CO$_2$ and H$_2$O clouds

We isolated 2 spectral categories of clouds in our dataset. A typical example is shown in Figure 3. The first category of clouds is characterized by a strong water ice absorption feature at 3 µm, frequently associated with water ice absorption features at 1.5 µm and 2 µm. The presence of these characteristics water ice features imply that these clouds are dominated by water ice. On the contrary, the second category of clouds shows no water ice features in their ratio spectrum (Figure 3a and b), even after performing averaging to enhance the signal to noise ratio. Two hypotheses can then explain these clouds: (1) they are composed of water ice, but due to their small particle size or near-IR optical depth, the water ice features are not detectable; (2) they are not primarily composed of water ice, which indicates a $CO_2$ ice composition if we make the reasonable assumption that $H_2O$ and $CO_2$ are the only condensable species.

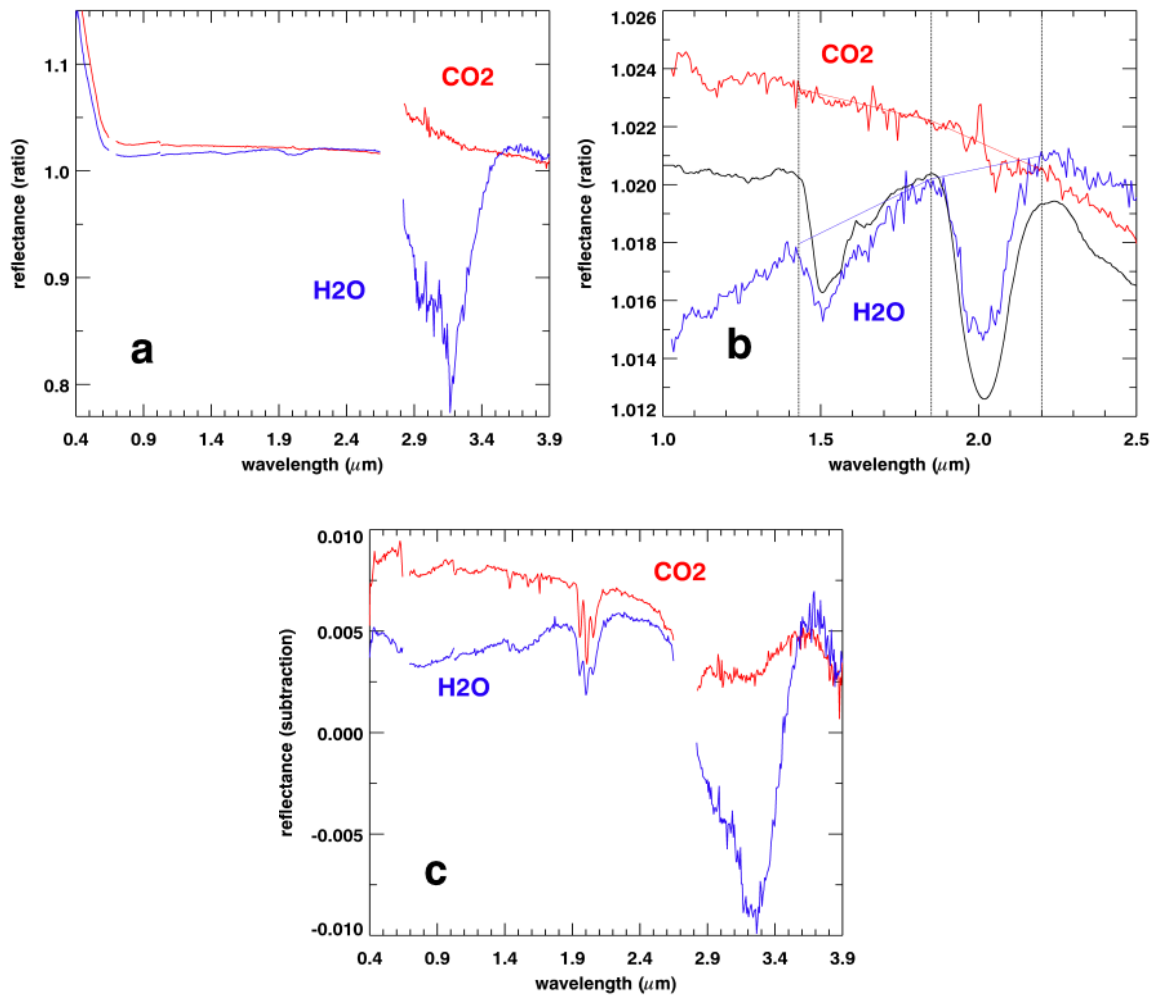

*Figure 3: Spectral properties of typical water ice (blue) and $CO_2$ ice (red) CRISM clouds. **(a)** and **(b)**: spectral ratios of the cloud divided by the adjacent cloud-free area. Water ice clouds have a strong water ice absorption feature at 3 µm **(a)** + frequent 1.5 and 2 µm water features **(b)**, while $CO_2$ clouds have no absorption features. A typical atmospheric water ice spectrum extracted from CRISM observations near Olympus Mons is shown **(b)** for comparison (black line). Increases in ratios for wavelengths < 0.7 µm ($H_2O$ and $CO_2$ clouds) and about 3 µm ($CO_2$ cloud) are due to the extremely low reflectance of the surface at these wavelengths (caused respectively by ferric oxides and adsorbed water) which dominates the ratio otherwise dominated by the cloud. **(c)**: spectral subtraction of the cloud minus the cloud-free area. The brightness level is linked with the optical depth (optical depths about 0.1 – 0.2 for reflectances of 5-10‰). The spectral slope is linked with the mean grain size of the clouds (typically 1 – 4 µm here). The interpretation of absorption features is less straightforward in*

*subtractions than in ratios (the large $CO_2$ gas absorption near 2 μm does not disappear; the $CO_2$ cloud shallow and broad feature near 3 μm is discussed in more detail in the text).*

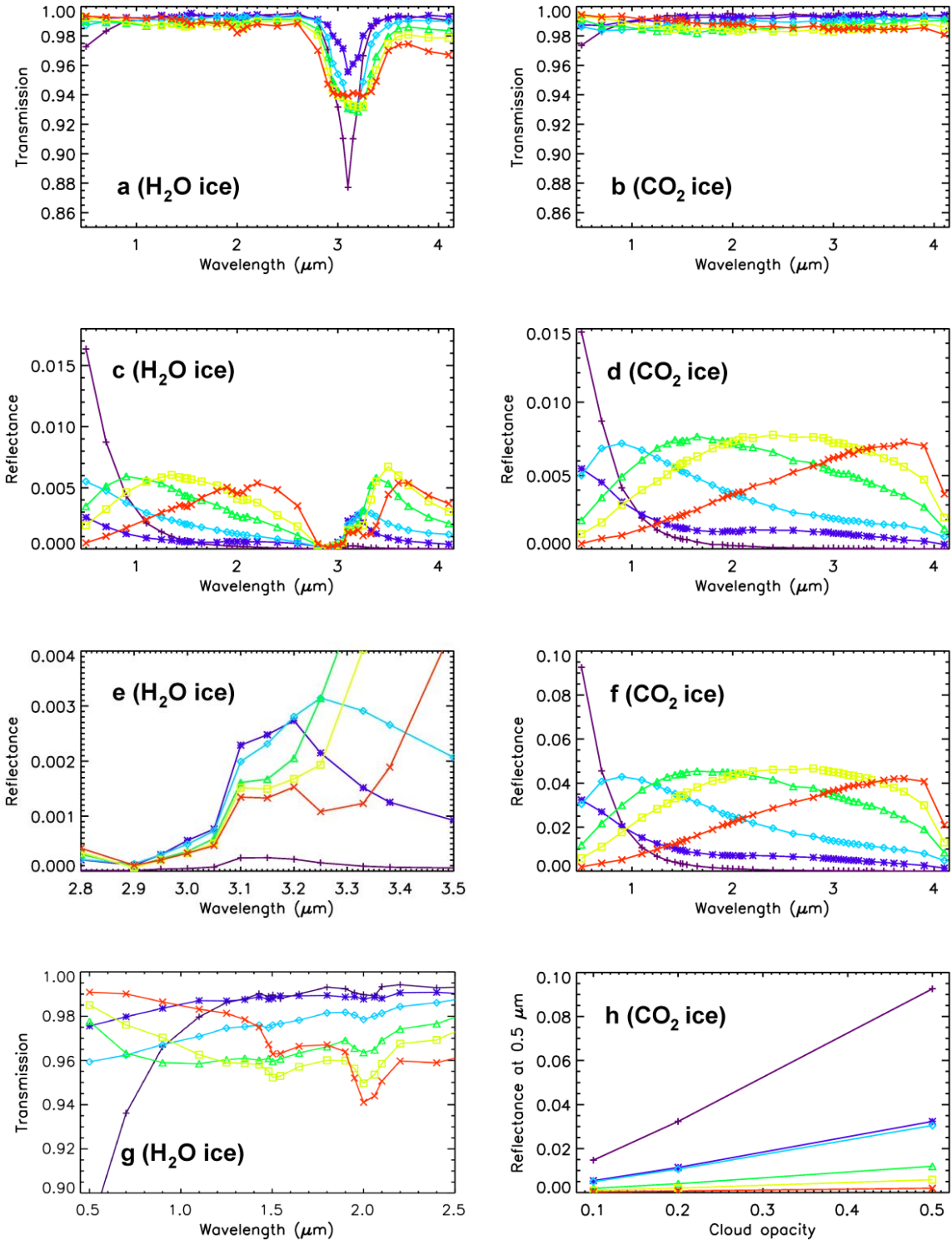

*Figure 4: Radiative transfer modeling of clouds spectral properties. **(a, c, e, g)** $H_2O$ ice. **(b, d, f, h)** $CO_2$ ice. **(a, b, g)** Transmission "$T_{Cloud}$". **(c, d, e, f, h)** Reflectance "$S_{Cloud}$" (for a solar zenith angle of 45°). Various effective radius are considered: purple crosses: $r_{eff}$=0.1 μm; blue stars: 0.5 μm; light blue diamonds: 1 μm; green triangles: 2 μm; yellow squares: 3 μm; red crosses: 6 μm. Calculations are performed for an optical depth of τ = 0.1, except for **(f, g)**: τ = 0.5 and **(h)**: τ = 0.1, 0.2 and 0.5. The*

*transmission of H₂O ice is characterized by a strong 3 µm feature **(a)** and 2 shallower bands at 1.5 µm and 2 µm **(g)**, while the transmission of CO$_2$ ice **(b)** is nearly wavelength-independent. Cloud scattering is strongly sensitive to the particle size, with a progressive shift of the scattering maximum from shorter to longer wavelength as the particle size increases **(c, d, f)**. The scattering of water ice clouds **(c, e)** also contains a complex 3 µm feature which results from the combination of a sharp 3.2 µm maximum (mainly visible for small grain sizes) and a wider 3 µm band (mainly visible for large grain sizes). Changing the optical depth within the 0.05 – 0.5 range mainly scale properties by a constant factor **(d, f, h)**.*

We performed Mie and radiative transfer modeling to assess the spectral properties of $H_2O$ and $CO_2$ ice clouds of various particle size and optical depth. We computed the single scattering parameters of various Gamma size distributions of ice particles using a Mie radiative transfer code (Mishchenko & Travis 1998). We used the $H_2O$ ice optical constants gathered by (Schmitt et al. 2004) for a 145 K temperatures, which are composed of values from (Grundy & Schmitt 1998; Trotta 1996) for wavelengths greater than 1 µm and values from (Warren 1984) for wavelengths less than 1 µm. For $CO_2$ ice, we used the $CO_2$ optical constants measured by (Schmitt et al. 1998) for a temperature of 15 K. While 145 K is a relevant temperature for martian atmospheric water ice (Smith 2004), 15 K is too low as mesospheric $CO_2$ clouds will typically have higher temperatures about 80 K (Montmessin et al. 2006). We compared our $CO_2$ ice constants with measurements by (Hansen 2005; Hansen 1997) acquired at 150K and found no significant differences within our wavelength range given our spectral resolution. A Monte-Carlo multiple scattering radiative transfer code (Vincendon et al. 2007) is then used to simulated the impact of clouds with various optical depths and mean grain sizes on solar radiance. Results are presented in Figure 4.

We explored a range of visible optical depths varying from 0.05 to 0.5 at 0.5 µm and 6 different particle size distributions ($r_{eff}$ = 0.1 µm and $v_{eff}$ = 0.05, $r_{eff}$ = 0.5 µm and $v_{eff}$ = 0.05, $r_{eff}$ = 1 µm and $v_{eff}$ = 0.1, $r_{eff}$ = 2 µm and $v_{eff}$ = 0.1, $r_{eff}$ = 3 µm and $v_{eff}$ = 0.1, $r_{eff}$ = 6 µm and $v_{eff}$ = 0.1). This range corresponds to detectable clouds using our approach (see section 2.a), as smaller cloud optical depths will not significantly scatter light at 0.5 or 1.3 µm (Figure 4). The spectral behavior of $CO_2$ clouds is mainly featureless: the transmission factor smoothly varies about 0.99 for an optical depth of 0.1 (Figure 4b). On the contrary, $H_2O$ clouds transmission contains a strong 3 µm absorption feature (Figure 4a), and two shallower 1.5 µm and 2 µm bands which detection requires a high signal to noise ratio for small-grained clouds (Figure 4g). Small-grained $CO_2$ clouds (0.1 µm) mainly scatter at shorter visible wavelengths, as we are close to the Rayleigh regime for this grain size (Figure 4d). Larger grain sizes significantly scatter at all near-IR wavelengths; the wavelength of the maximum is roughly equal to the mean grain size (Figure 4d). The scattering pattern of $H_2O$ ice is the combination of a similar wavelength slope and a strong absorption (large grains) or emission (small grains) feature near 3 µm (Figure 4c, e). This complex spectral behavior results from the increase of the attenuation coefficient of water ice from 0.01 at 2.8 µm / 3.6 µm to ≈ 1 at 3.1 µm (Warren 1984; Grundy & Schmitt 1998). As the attenuation coefficient increases near 3.1 µm, the refractive index also increases from 1.3 to 1.8. For small particles, this attribute results in a strong increase of the quantity of light scattered by water ice grains at 3 µm to 3.6 µm compared to surrounding wavelengths. A similar effect is observed for $CO_2$ ice at wavelengths about 4.3 µm (Montmessin et al. 2007). For both $H_2O$ and $CO_2$ clouds, changing the optical depth between 0.05 and 0.5 mainly results in scaling figures by a constant factor (Figure 4d, f, h).

To summarize, water ice clouds are always characterized by a diagnostic 3 µm water ice feature, frequently accompanied by 1.5 µm and 2 µm features (see also (Langevin et al. 2007)). The 3 µm feature is already several % deep in the transmission spectrum of thin clouds with normal optical depth of 0.05 – 0.1 corresponding to our limit of detection. All clouds detected using our approach (section 2.a) will therefore show diagnostic features of water ice in their transmission spectra (~ratio) if they are composed of water ice. On the contrary, $CO_2$ ice is characterized by a constant transmission spectrum. As a consequence, the spectral properties of clouds near 1.5, 2 and 3 µm are diagnostic of their $H_2O$ versus $CO_2$ composition: clouds that lack water ice features in their ratio spectrum are $CO_2$ clouds (Figure 3). On this basis, we have inspected the spectral properties of all CRISM cloudy observations and we have classified clouds as $H_2O$ (Table 2) or $CO_2$ (Table 1) accordingly. We can see on Figure 3c that a shallow but wide feature near 3 µm is observed in the "subtraction" spectrum derived for $CO_2$ clouds. This feature is not expected for $CO_2$ ice clouds scattered reflectance (Figure 4). The origin of this feature is unclear: as explained in section 2.b, the "subtraction spectrum" is only a first order estimate of the cloud scattered reflectance, it still contains part of the surface reflectance; as the surface of Mars is characterized by a strong wide 3 µm absorption band, the shallow and wide 3 µm band observed in the subtraction spectrum could be a remnant of this surface feature. However, our modeling indicates that $T_{Cloud}$ is slightly lower than 1 (Figure 4b), which would results in a weak contribution of an upturned surface 3 µm feature, while we observe an actual 3 µm feature. A change in the assumptions regarding the single scattering properties of $CO_2$ ice particles could result in $T_{Cloud} > 1$, which would explain the stronger water ice feature observed above the cloud. Alternatively, this feature could betray the presence of a few water ice particles within the $CO_2$ cloud, or even within $CO_2$ ice grains as water ice grains could act as condensation nuclei for $CO_2$ ice.

### d. Observational biases

We provide in this section information about the various observational biases associated with CRISM observations of clouds. This will help comparing cloud properties retrieved with CRISM, such as apparent morphology or spatial distribution, with information provided by previous studies based on other instruments.

First, at high resolution, CRISM observes a given surface scene of about 11x11 km with a varying emergence angle: the pointing changes to compensate for the spacecraft motion in order to increase the integration time. The orbit of MRO is nearly polar: the emergence (emission angle) varies from about -35° in the south part of a given image to 0° (nadir viewing) in the center of the image to +35° in the north part. As the field of view of CRISM remains constant, this procedure creates the hourglass shape seen in CRISM image once projected on the ground (Murchie et al. 2007). As a consequence, at 75 km altitude – the typical altitude of $CO_2$ clouds (Scholten et al. 2010; Määttänen et al. 2010) – a surface of 11x116 km is intercepted in the field of view: high altitude clouds are therefore compressed by a factor of about 10 in the nearly north-south direction of the MRO orbit, without being changed in the east-west direction. This effect is illustrated in Figure 5: a $CO_2$ cloud with NW-SE stripes observed by HRSC is idealized as linear cloud features at 75 km altitude and projected in the CRISM observing geometry to estimate biases resulting from this specific CRISM viewing geometry. The altitude of known $CO_2$ ice clouds typically ranges from 50 to 100 km (Määttänen et al. 2010; Scholten et al. 2010; Montmessin et al. 2007; Montmessin et al. 2006; Clancy et al. 2007). This altitude range corresponds to North-South compression factors between 7 and 14.

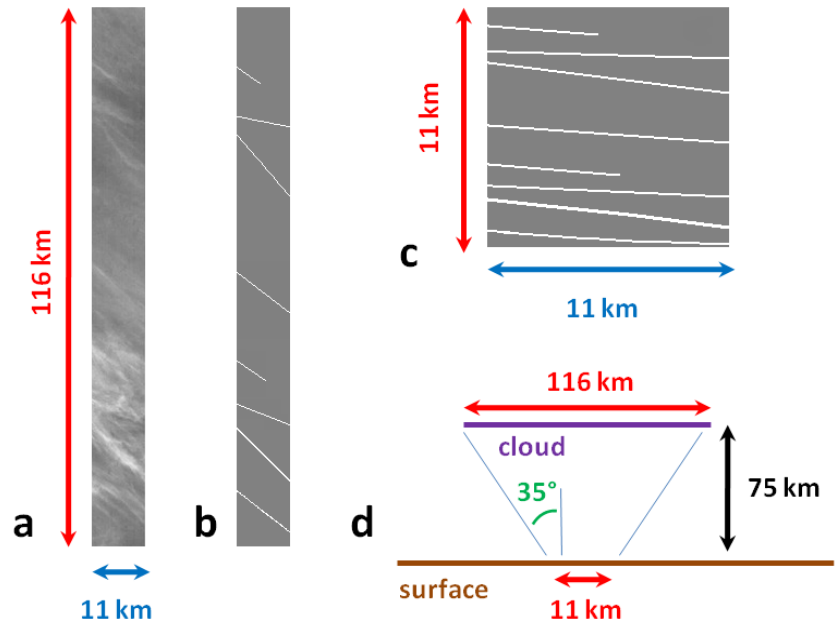

*Figure 5: Geometric distortion in CRISM $CO_2$ cloud images. (a) 11km x 116 km part of an HRSC image containing a $CO_2$ cloud, from (Scholten et al. 2010). (b) Scheme of this image. (c) 11km x 11km image of this scheme as it will be observed by CRISM for a cloud altitude of 75 km. (d) Scheme of CRISM viewing geometry. High altitude clouds are compressed in CRISM images due to the ±35° emergence angle variations implemented in the nearly north-south direction during targeted CRISM images acquisition.*

Secondly, the CRISM dataset tends to be biased toward surface sites that have been intensively studied for their mineralogical diversity, such as Mawrth Vallis, Terra Meridiani, Valles Marineris and Nili Fossae. We can see in Figure 6 that the distribution of CRISM observations is to first order well-scattered through the whole Martian planisphere. However, we can see that observational biases are responsible for small shape details of $CO_2$ clouds clusters.

Thirdly, there are also some observational biases which depend on the solar longitude:

- The distribution of CRISM observations as a function of $L_s$ is slightly irregular, notably due to coverage gaps related to MRO safe mode or instrumental problem occurrences in 2009 and 2010. However, the distribution of equatorial $CO_2$ cloud occurrence is not correlated to these variations (Figure 6c).
- A very limited number of observations have been acquired by CRISM at mid northern latitudes (40°N - 60°N) during northern autumn (Figure 6b). This will prevent us from doing comparison with cloud detections previously reported in this area (Figure 7).

Finally, as our cloud detection method is based on the visual observation of cloudy features in CRISM high resolution images, we will not detect clouds that are uniform over large areas greater than the extent of CRISM images (i.e., larger than 11 km in the east-west direction and up to 110 km in the north-south direction; see Figure 5).

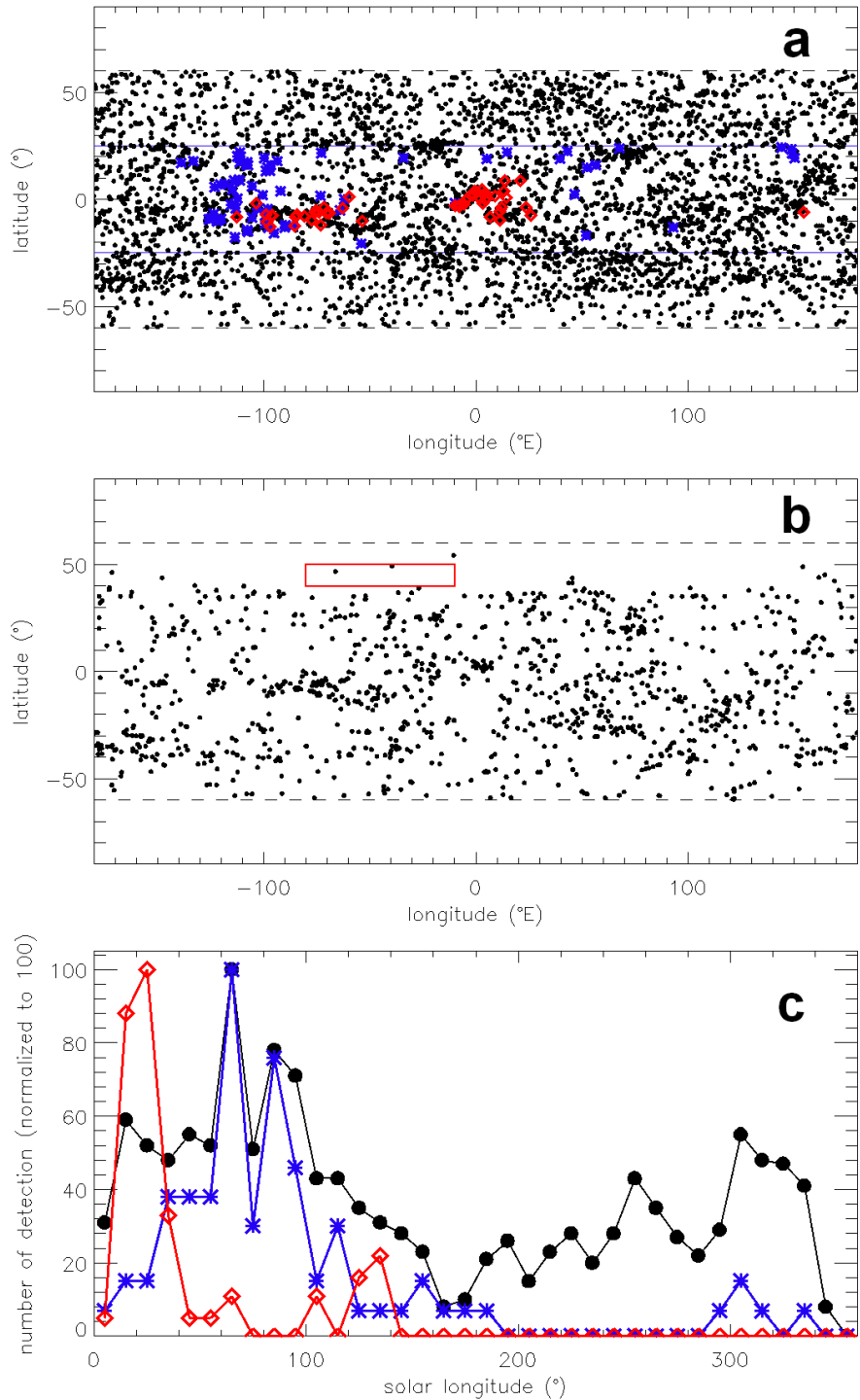

*Figure 6: Observational biases in the spatial/seasonal distributions of cloud detections. **(a)** CRISM $CO_2$ cloud observations (red diamonds) are compared to CRISM observations acquired between $L_S$ 0 and $L_S$ 140° (black dots) and to equatorial (± 25° latitude) $H_2O$ clouds (blue stars). **(b)** CRISM observations acquired in the $L_S$ 215°-275° range (black dots) compared to the location (red square) of high altitude or $CO_2$ clouds observed in this $L_S$ range (Figure 7). **(c)** CRISM equatorial (±25° latitude) observations compared to CRISM $CO_2$ (red diamonds) and $H_2O$ (blue stars) clouds.*

## 3. Results and discussion

### a. Spatial distribution

We have identified 54 observations of $CO_2$ clouds in the CRISM dataset (Table 1). All of these observations are located in the [13°S, 9°N] latitude range (we have searched for $CO_2$ clouds between 60°S and 60°N). The spatial and time distributions of these clouds are compared to the distribution of $H_2O$ clouds observed by CRISM in the [25°S, 25°N] range in Figure 6. We can see that clouds classified as $H_2O$ or $CO_2$ on the basis of their spectral properties have very distinct spatial and seasonal patterns.

We compare the spatial and $L_S$ distributions of CRISM $CO_2$ clouds to previously published detections of $CO_2$ or high altitude clouds (> 60 km) in Figure 7, Figure 8 and Figure 9. OMEGA observations refer to the 50 Mars Express orbits previously published for Mars Years (MY) 27 to 29 (Määttänen et al. 2010), plus 17 new Mars Express orbits for MY 30 (Table 3) where we have detected $CO_2$ clouds using the same approach as originally developed by (Montmessin et al. 2007). Thirteen of these seventeen new detections are scattered (in a relatively homogeneous manner) between $L_S$ 10° and $L_S$ 80°, three have been obtained at $L_S$ 115-123° and one at $L_S$ 151°. A given Mars Express orbit can contain widespread $CO_2$ clouds extending over several tens of latitude degrees: we have sampled these OMEGA clouds every 5° latitude and obtain 150 OMEGA cloud occurrences that can be compared to CRISM observations. High altitudes clouds refer to TES/MOC "MEM" clouds (Clancy et al. 2007), HRSC "$CO_2$" clouds (Scholten et al. 2010; Määttänen et al. 2010), SPICAM "$CO_2$" clouds (Montmessin et al. 2006) and THEMIS high altitude clouds (> 60 km) (McConnochie et al. 2010).

Table 3: list and properties of the new OMEGA observations with $CO_2$ clouds.

| Observation | year | latitude (°) | longitude (°E) | Ls (°) |
|---|---|---|---|---|
| 7536 | 2009 | [0 ; 4] | 7.7 | 11.0 |
| 7537 | 2009 | [-5 ; 1] | 268.0 | 11.1 |
| 7554 | 2009 | [-2 ; 1] | 353.6 | 13.5 |
| 7604 | 2009 | [-3 ; 4] | 351.0 | 20.3 |
| 7679 | 2009 | [-2 ; 6] | 345.5 | 30.3 |
| 7686 | 2009 | [-7 ; -4] | 359.0 | 31.2 |
| 7768 | 2010 | 4.5 | 8.5 | 41.9 |
| 7875 | 2010 | 4.0 | 3.5 | 55.5 |
| 7907 | 2010 | [-2 ; 1] | 341.0 | 59.6 |
| 7914 | 2010 | 5.0 | 356.0 | 60.5 |
| 7953 | 2010 | [5 ; 9] | 343.0 | 65.4 |
| 8020 | 2010 | [-2 ; 18] | 335.5 | 74.0 |
| 8048 | 2010 | [2 ; 5] | 345.5 | 77.5 |
| 8335 | 2010 | -8.0 | 357.9 | 114.8 |
| 8371 | 2010 | -16.0 | 275.1 | 119.7 |
| 8395 | 2010 | [-10 ; -7] | 314.3 | 123.0 |
| 8592 | 2010 | -1.0 | 349.0 | 151.2 |

CRISM $CO_2$ clouds are exclusively observed near Valles Marineris (between 113°W and 53°W) and near Terra Meridiani (between 10°W and 26°E), with the exception of one detection at 155°E. All CRISM detections except the 155°E one fall within previously known areas of $CO_2$ cloud activity as derived from the composition-based OMEGA $CO_2$ cloud detections. The 155°E detection occur in an area of known sporadic $CO_2$ cloud activity (one OMEGA observation at 120°E and two HRSC (Scholten et al. 2010) "high altitude" cloud observed in the neighborhood). The OMEGA+CRISM pattern of $CO_2$ cloud activity is generally in agreement with TES/MOC and HRSC observations of mesospheric clouds, except for a few TES/MOC observations near 85°W and 10°N (Figure 7). We do not observe with daytime CRISM and OMEGA observations any cloud with a $CO_2$ spectroscopic composition in the latitude range 30-40°S where SPICAM has observed 2 nighttime high altitude bright clouds referred as "$CO_2$" clouds. This latitude range corresponds to an area of frequent cold "$CO_2$ ice compatible" mesospheric temperatures at night based upon SPICAM mesospheric temperature retrievals (Forget et al. 2009). These two SPICAM clouds are also very distinct from other clouds in latitude/season and longitude/season diagrams (Figure 9). No $CO_2$ clouds are seen by CRISM at mid latitudes, while $CO_2$ (OMEGA) or high altitude (THEMIS) clouds are reported between 40°N and 50°N for $L_S$ in the [215° - 275°] range (Figure 7). As discussed in section 2.d, only a few CRISM observations have been acquired for these latitudes at appropriate $L_S$. Hence, this mismatch is not significant.

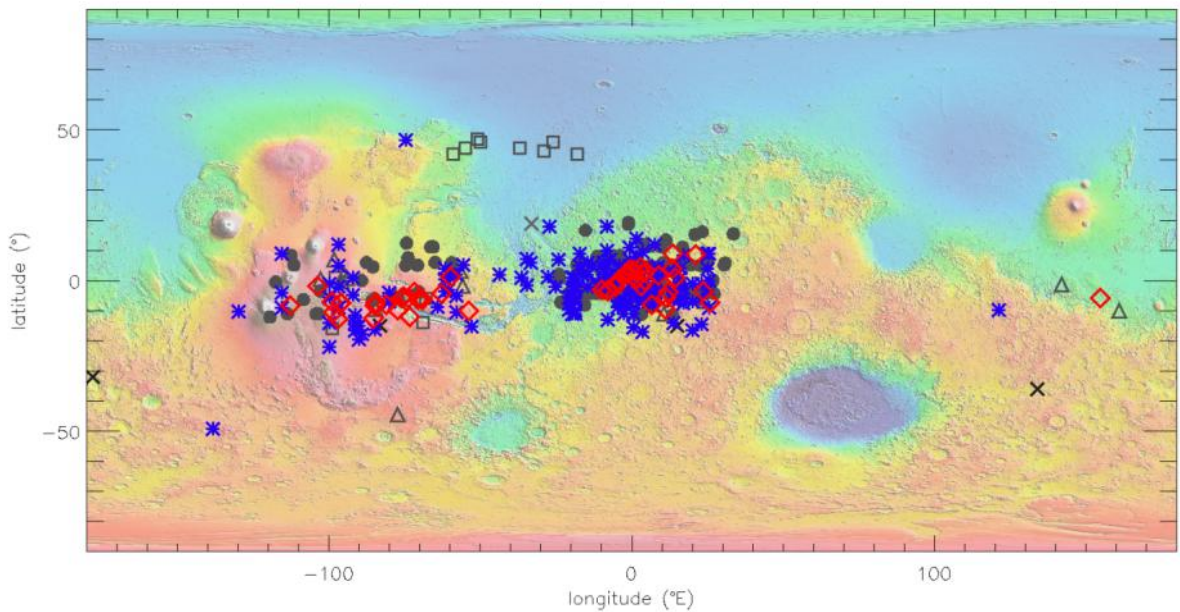

*Figure 7 : CRISM (red squares) and OMEGA (blue stars) spectroscopic identification of $CO_2$ clouds compared to mesospheric (≥ 60 km) clouds observed by TES/MOC (dots), HRSC (triangles), SPICAM (dark crosses), THEMIS (squares) and Pathfinder (light star). CRISM detections, from a survey in the 60°S-60°N range, are all located within ± 15° latitude about the equator, near Valles Marineris or Meridiani except for a detection at 155°E. CRISM detections are in very good agreement with OMEGA $CO_2$ cloud detections. CRISM observations are also consistent with most TES/MOC and HRSC mesospheric clouds, as well as with SPICAM and THEMIS equatorial mesospheric clouds.*

### b. Time distribution

All CRISM $CO_2$ clouds are observed between $L_S$ 0° and $L_S$ 138° (Figure 8), with a main maximum at $L_S$ 20° and a secondary maximum at $L_S$ 130°. Between $L_S$ 0° and $L_S$ 40° (main maximum), about 30% of CRISM equatorial (±10° latitude) observations near Valles Marineris and Terra Meridiani contain $CO_2$ clouds. CRISM $L_S$ distribution is in good agreement with OMEGA detections, as well as with most TES/MOC and HRSC mesospheric cloud occurrences. There are a few differences between CRISM and OMEGA after $L_S$ 40°: the $L_S$ 20° peak ends at $L_S$ 40° for CRISM versus $L_S$ 60° for OMEGA; the frequency of clouds increases from $L_S$ 100° to $L_S$ 135° in CRISM data while it decreases in OMEGA data. However, these differences may be biased by the low number (12) of clouds observed by CRISM between $L_S$ 40° and $L_S$ 140°. A distinct peak activity in mesospheric clouds is reported by TES/MOC between Ls 140° and Ls 166° (Figure 8, Figure 9), which is not observed by either OMEGA or CRISM (OMEGA clouds are observed up to Ls 135°, + one cloud at Ls 151°). The pathfinder cloud (located at 19°N and 33°W and observed at night before dawn), interpreted as a $CO_2$ cloud (Clancy & Sandor 1998), also occur at later Ls of 162°. We note that the Martian years of OMEGA and CRISM observations differ from that of TES/MOC and Pathfinder observations. As explained in section 2.d, northern mid latitudes are infrequently observed by CRISM during northern fall, which prevents comparisons with the THEMIS and OMEGA cloud detections for the corresponding $L_S$ range ($L_S$ > 200°, see Figure 9).

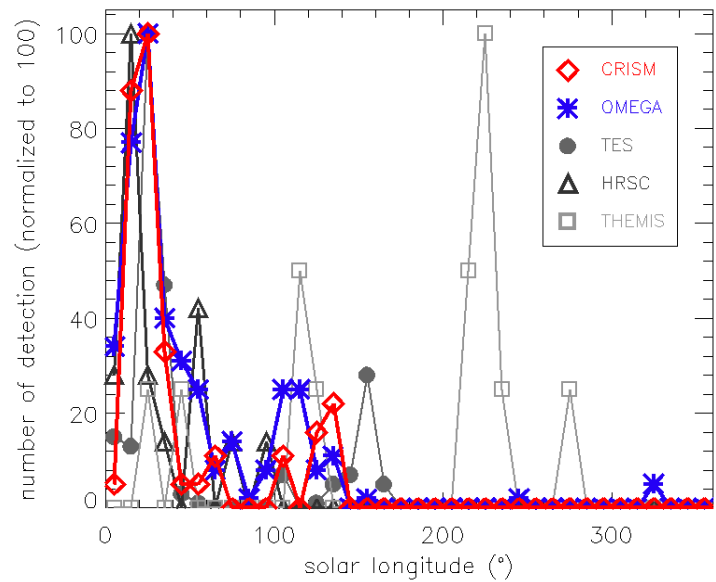

*Figure 8: Distribution of CRISM (red squares) and OMEGA (blue stars) spectroscopic identification of $CO_2$ clouds as a function of $L_S$, compared to mesospheric (≥ 60 km) clouds observed by TES/MOC (dots), HRSC (triangles) and THEMIS (squares).*

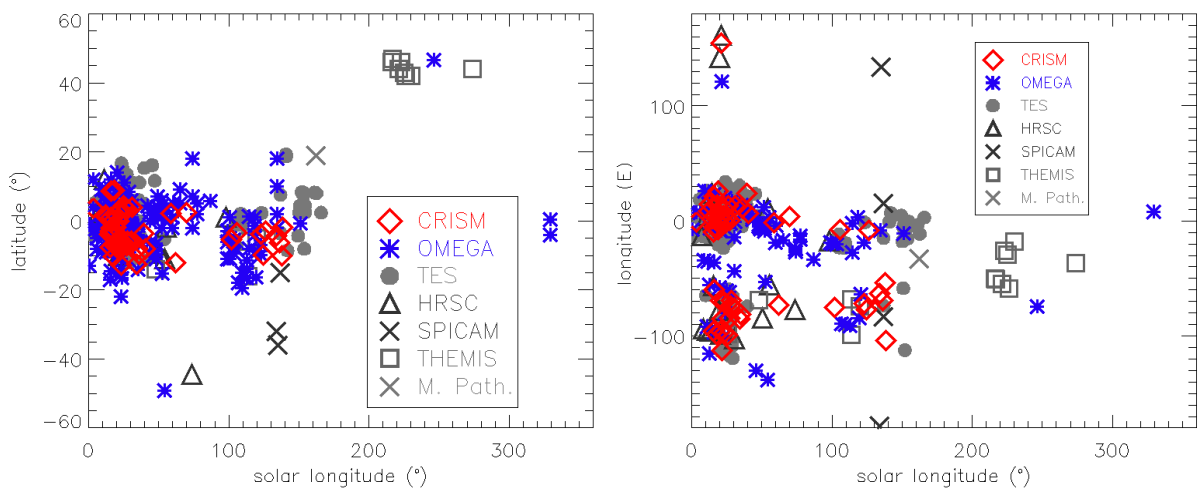

*Figure 9: CRISM and OMEGA spectroscopic identification of $CO_2$ clouds compared to mesospheric (≥ 60 km) clouds observed by TES, HRSC, SPICAM, THEMIS and Pathfinder. Latitude (left) or longitude (right) versus solar longitude is represented.*

### c. Composition of high altitude clouds from OMEGA limb

This extensive comparison of various datasets highlights a few significant differences between spectroscopy-based identification of $CO_2$ clouds (OMEGA and CRISM) and mesospheric clouds interpreted as probable $CO_2$ clouds (TES/MOC, SPICAM, THEMIS, HRSC). It is difficult to draw conclusions from comparison only, as many factors could contribute to these discrepancies:

- Interannual variability: OMEGA and CRISM observations barely overlap with TES/MOC observations (MY 26-29 for CRISM-OMEGA versus MY24-26 for TES). They however overlap with SPICAM, HRSC and THEMIS observations.
- Local time: OMEGA observations are restricted to daytime and CRISM observes about 15.00 local time. On the contrary, SPICAM limb profiles are acquired during nighttime, THEMIS observations of mid latitudes clouds was obtained at twilight and the Pathfinder cloud was observed before sunrise. TES/MOC observations of mesospheric clouds have however also been acquired during daytime.
- Viewing geometry: thin clouds could be undetectable in nadir viewing geometries (OMEGA, CRISM) while seen in limb viewing geometries (TES/MOC, SPICAM). However, THEMIS also observes in nadir and the 3 OMEGA limb detections of $CO_2$ clouds reported so far are within OMEGA nadir clouds areas (Gondet et al. 2008).
- "$CO_2$ composition" detection limit of both OMEGA and CRISM. A very distinct technique has however been used to detect $CO_2$ clouds with CRISM compared to OMEGA.
- $H_2O$ clouds could be misinterpreted as $CO_2$ clouds when no composition measurements are available. However, $H_2O$ clouds are usually found at altitudes up to 60 km only (Heavens et al. 2011).

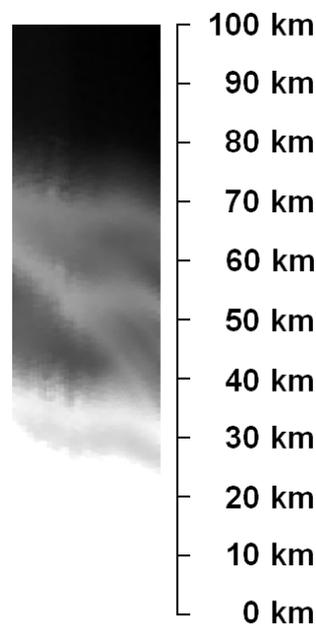

*Figure 10: OMEGA limb observation # 1322_2. This observation has been obtained above longitude 18°E and latitudes 30-37°S for $L_S$ 150° on January 27 of 2005. The local time is 13.00. The flux received by OMEGA at 0.5 µm while looking at the atmosphere above the limb of the planet is represented as a function of altitude. Bright layers extending above 70 km are observed.*

To bring new constraints about this last hypothesis, we have analyzed OMEGA limb observations acquired at time and places where discrepancies are noticed. One OMEGA limb observation showing high altitude bright layers extending up to 75-80 km has been obtained at $L_s$ 150° above latitudes 30-37°S for a longitude of 18°E (Figure 10). This observation can be compared to the high altitude clouds detected by SPICAM at the same latitudes a little earlier in the season ($L_s$ 135°) and by TES/MOC at the same $L_s$ a little closer to the equator (< 20°S).

At visible and shorter near-IR wavelengths, bright atmospheric layers are observed at altitude extending up to 75-80 km (Figure 11). The spectral properties of these high altitude aerosols are diagnostic of $H_2O$ ice (Figure 12): as discussed in the modeling section (2.c), small-grained (0.1 - 0.5 µm) water ice aerosols present a scattering maximum at 3.1 – 3.3 µm. On the contrary, no $CO_2$ ice signatures are observed at 4.26 µm. The observation of water ice clouds at such high altitudes seems consistent with preliminary results derived from MCS data (Kass et al. 2011), and with haze top altitude inferred from SPICAM stellar occultations (Montmessin et al. 2006).

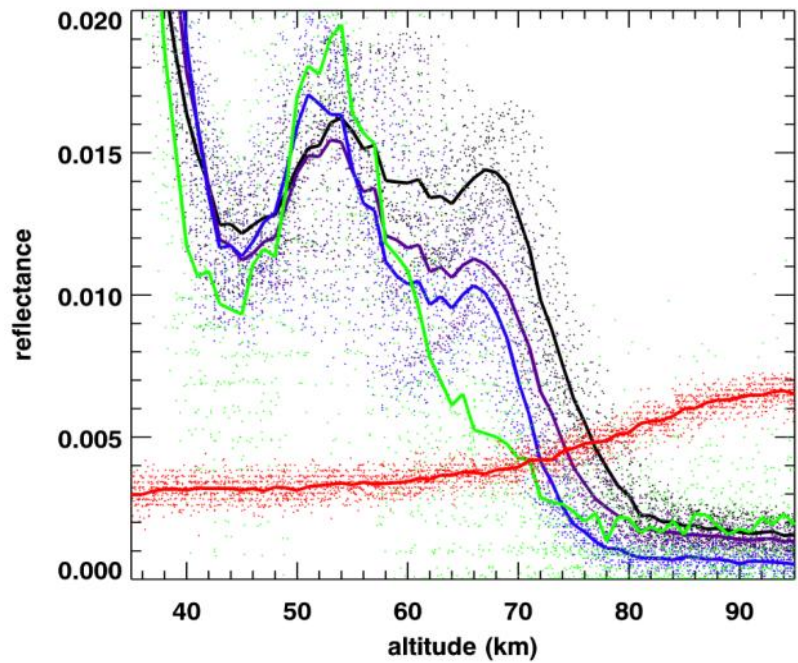

*Figure 11: Variations of the reflectance received by OMEGA as a function of the altitude for 5 different wavelengths: black, 0.5 µm; purple, 0.75 µm; blue, 1.1 µm; green, 3.1 µm; red, 4.26 µm. Reflectances are scaled by a constant factor for clarity (respectively 0.5, 0.5, 1, 3.3 and 0.1). Bright layers at visible (black and purple) and near-IR (blue and green) wavelengths extend at high altitudes up to 75-80 km. The detectable position of the altitude maximum decreases with wavelength, from about 80 km to 75 km: this could result from the expected decrease of the mean particle size with altitude. These layers present the diagnostic 3.1 µm (green) water ice maximum (see Figure 12). On the contrary, no "4.26 µm" $CO_2$ ice feature (Montmessin et al. 2007) is associated with these layers: the reflectance at 4.26 µm (red) only increase at high altitude (> 80 km) due to the non-LTE fluorescent emission of $CO_2$ gas (Drossart et al. 2005).*

This observation demonstrates that water ice aerosols can be found at very high altitudes where detached aerosols layers have been previously interpreted as $CO_2$ ice layers. In particular, the 2 "detached layers" observed by SPICAM UV sensor at similar latitudes (~ 35°S) and solar longitude (~ 135°) at 75-95 km (Montmessin et al. 2006) could in fact correspond to water ice aerosols, and not $CO_2$ ice aerosols as claimed in the title of this publication. We can see on Figure 11 that the top altitude of the water ice layer slightly increases as wavelength decreases: this is consistent with a vertical gradient of particle size, with smaller particles at higher altitude. As a consequence, at UV wavelengths, the top of this water ice layer will be observed at even higher altitudes. This is also consistent with the fact that water ice clouds are usually observed at slightly lower altitudes at far-IR wavelengths (Heavens et al. 2011). Moreover, SPICAM observes at night while OMEGA observes during the day, and water ice clouds are expected to be observed at even higher altitude at night (Heavens et al. 2010). The simultaneous presence of temperature below the $CO_2$ frost point was a major reason to interpret SPICAM clouds as $CO_2$ ice clouds. However, the cold, $CO_2$ ice compatible temperatures are systematically found by SPICAM at altitudes above the top of the detached layers (Montmessin et al. 2006), and numerous – about 8% – SPICAM observations show such subfreezing temperatures, without detectable nearby ice particles for a large majority of them (Montmessin et al. 2011). Particles size also matches, as both OMEGA high altitude water ice clouds and SPICAM clouds correspond to small, submicron grains. On the contrary, $CO_2$ clouds are usually larger-grained (see section 2.d). As a consequence, we believe that $H_2O$ ice is an interpretation at least as valid as $CO_2$ ice for these high altitude bright layers that are observed in areas where OMEGA and CRISM do not observe $CO_2$ clouds, but where OMEGA does observe mesospheric $H_2O$ clouds. Similarly, the peak in equatorial mesospheric cloud activity observed by TES/MOC in late northern summer could partly result from these high altitude late northern summer water ice clouds observed at low latitudes.

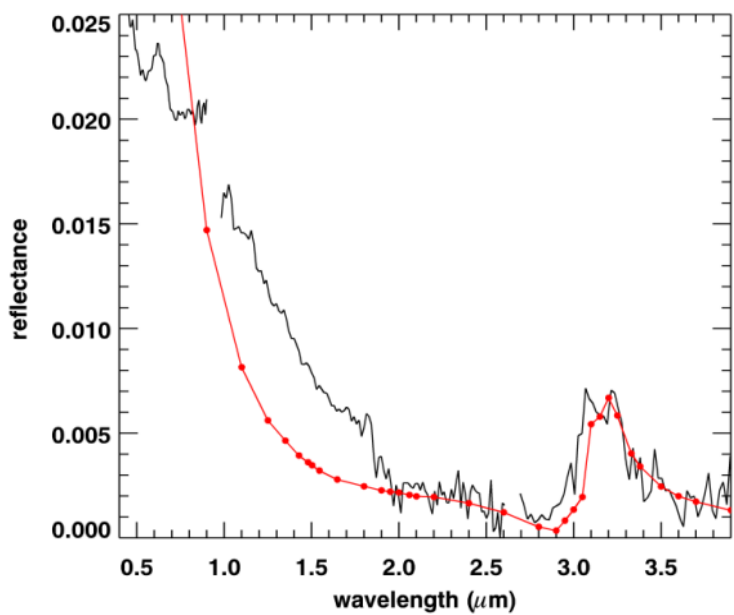

*Figure 12: OMEGA spectrum (black) of the high altitude aerosols layer of limb # 1322 (Figure 10). This particular spectrum corresponds to an altitude of 65 km. The spectral properties of the particles are characterized by a rapid decrease of the scattering efficiency from 0.5 to 1.5 µm, indicatives of submicron particles, and by a scattering maximum at 3.1 – 3.3 µm diagnostic of submicron water ice grains. A model spectrum extracted from Figure 4c is shown for comparison (red): it corresponds to an average of particle size 0.1 µm and 0.5 µm.*

### d. Optical depth and particle size estimate

We have derived the mean optical depth and particle size of a sample of CRISM $CO_2$ clouds using the modeling approach detailed in section 2.b. We show an example of this optical depth and particle size retrieval in Figure 13. Typically, $CO_2$ clouds particle size as derived from CRISM observations range from 0.5 µm to 2 µm. This compares very well with particle size estimate based on OMEGA data (Montmessin et al. 2007; Määttänen et al. 2010). Visible optical depths (0.5 µm) are typically of 0.2 (Figure 13), in agreement with (Montmessin et al. 2007), and present spatial variability within a given cloud filament, similarly to what is shown by (Määttänen et al. 2010). Normal optical depth retrieval based on limb observations (Clancy et al. 2007; Montmessin et al. 2006) where several order of magnitude smaller. However, these retrievals assume a spatially homogeneous and widespread cloud, which is not relevant as $CO_2$ clouds are localized with small scale structures.

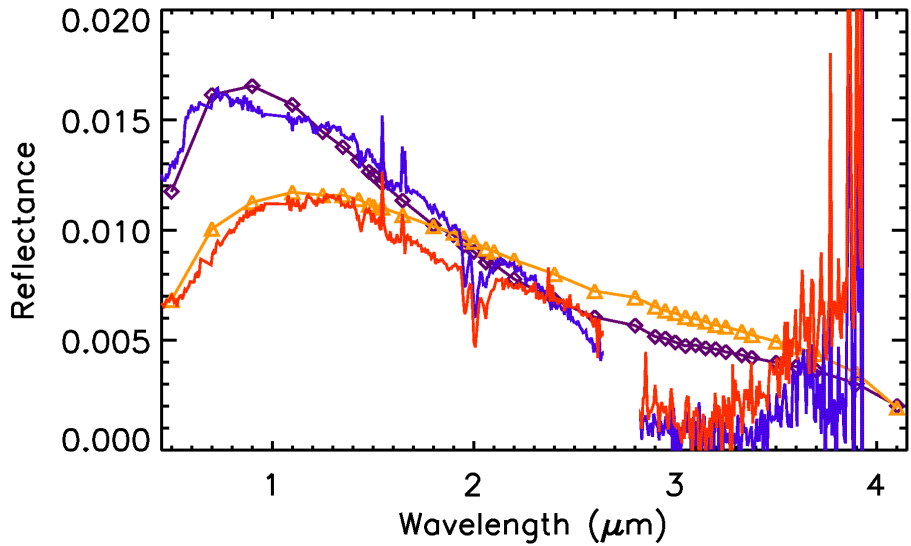

*Figure 13: CRISM $CO_2$ cloud particle size estimate. The scattered component of a CRISM $CO_2$ cloud (see Table 1) is extracted (red and blue lines) and compared to model results presented in Figure 4c (orange triangles and purple diamonds). The scattered component is derived by subtracting the surface spectrum to the cloud spectrum (see section 2.b). We have also taken into account a typical cloud transmission of 0.99 (Figure 4b). The highest uncertainty is related to the choice of the nearby surface spectrum: the two (red and blue) observed spectra correspond to 2 extreme possible choices in surface spectrum. Then, we have looked for the best fit of the model to these observations by adjusting the optical depth at 0.5 µm and the mean particle size. The resulting values are $r_{eff}$ = 1.0 µm and τ = 0.22 for the blue spectrum and $r_{eff}$ = 1.4 µm and τ = 0.17 for the red spectrum.*

### e. Morphology

The morphology of equatorial $CO_2$ clouds has been first described as mostly "cumuliform" on the basis of a limited sample of OMEGA observations, suggesting (with other theoretical considerations) a potential convective formation mechanism (Montmessin et al. 2007). A revised analysis of an extended set of OMEGA data had lead to the conclusion that only a minor fraction of OMEGA observation – mostly obtained during morning hours - can be classified as cumuliform, while most OMEGA observations (mostly obtained during mid-afternoon) show either cirrus-type morphologies or undefined morphologies. A lower limit of the cumuliform percentage has been estimated to 15% (Määttänen et al. 2010). This reanalysis was supported by the higher spatial resolution HRSC observations which showed that "the prevailing morphologies are ripple-like or filamentary forms" (Scholten et al. 2010), and that a given cloud can appear "clumpy" at low spatial resolution while "high spatial resolution reveals a mainly filamentary interior structure of the cloud" (Scholten et al.

2010). The orientation of the filaments in HRSC data is "roughly E-W, sometimes NW-SE". The HRSC observations reported in these publications are limited in local time (afternoon hours) and $L_S$ range (0-36°), while "the round-shaped OMEGA clouds have been observed in the morning except for one early afternoon cases in the northern mid-latitudes" (Määttänen et al. 2010). Further observations were therefore required at that time to determine to which extent the results derived from the analysis of HRSC data can be or not extended to all $CO_2$ clouds.

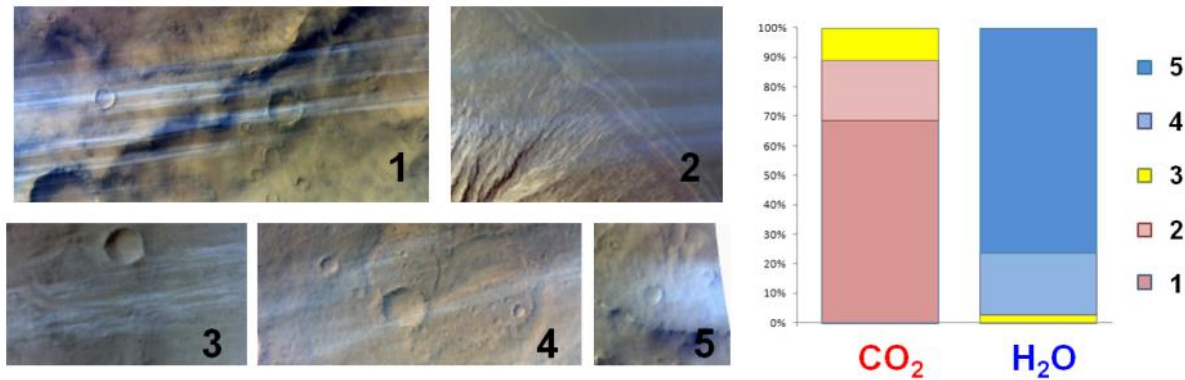

*Figure 14: Classification of the CRISM $CO_2$ and $H_2O$ clouds listed in Table 1 and Table 2 as a function of the apparent morphology (CRISM images contain geometric distortion, see Figure 5). We have divided our cloud sample in 5 typical apparent morphologies: straight filaments (1), blurred filaments (2), wavy filaments (3), elongated blurred areas (4) and rounded blurred areas (5). $CO_2$ and $H_2O$ clouds belong to distinct apparent morphology types (respectively 1-2 and 4-5 for most observations).*

We have classified the apparent CRISM cloud morphologies in Figure 14. There is a strong correlation between apparent morphology and composition: all $CO_2$ clouds are characterized by more or less straight filaments, while $H_2O$ clouds never show "straight filament" apparent morphologies but more often present blurred substructures (Figure 14). The apparent morphology of the clouds is impacted by cloud altitude due to the specific CRISM viewing geometry (Figure 5). However, $H_2O$ and $CO_2$ daytime cloud altitude distributions overlap: 0 – 80 km and 50 – 100 km respectively. As a consequence, this dichotomy also reflects a variation of the morphology depending on composition. We have compared a pair of simultaneous observations of a $CO_2$ cloud area by OMEGA and CRISM in Figure 15. This comparison show how the spatial resolution can impact our apprehension of $CO_2$ cloud morphology: while the OMEGA observation would have been classified as "cumuliform" or "undefined" (Määttänen et al. 2010), CRISM clearly show fine filament-like structures in this cloud. To conclude, morphologies in CRISM afternoon observations are consistent with HRSC results (Scholten et al. 2010), with clouds composed of filaments "roughly E–W" oriented. This extends the result of (Scholten et al. 2010) outside the $L_S$ 0-36° range as CRISM equatorial $CO_2$ cloud observations extend from $L_S$ 0° to 140°.

CRISM observations are restricted to afternoon hours, while most of the "round-shaped" OMEGA clouds classified as cumuliform are observed at morning hours (Määttänen et al. 2010). We have looked for additional observational constraints corresponding to morning hours. In Figure 16, we compare a CRISM afternoon $CO_2$ cloud to the pre-dawn cloud observes from the surface by Pathfinder and interpreted as a probable $CO_2$ cloud (Clancy & Sandor 1998). This cloud is described by these authors as a "bluish wave cloud". While taken in different conditions, Pathfinder (morning, from the surface) and CRISM (afternoon, from orbit) slanted observations show a remarkable agreement in terms of apparent morphology. Although there is no direct spectroscopic evidence for the Pathfinder cloud to be composed of $CO_2$ ice, this observation may indicate that morning $CO_2$ clouds are also composed of filaments. Similarly, the pathfinder cloud apparent morphology has been shown to match THEMIS equatorial high altitude clouds morphology (McConnochie et al. 2010). In Figure 17 we present a pair of HRSC/OMEGA observation of the same morning $CO_2$ cloud. The cloud, as seen with the low spatial resolution typical of OMEGA morning observations of $CO_2$ clouds, appear as a "rounded and clumpy, cumuliform cloud". However, the exact same cloud observed at the same time by HRSC, also onboard Mars Express, is similar to HRSC afternoon clouds with "cirrus-type" morphologies composed of filaments. As a consequence, our analysis does not support the conclusions of (Montmessin et al. 2007) and (Määttänen et al. 2010) that OMEGA observations reveal a cumuliform morphology for at least 15% of the clouds (mostly morning clouds): OMEGA data cannot be systematically used for cloud morphology interpretation due to its varying spatial resolution.

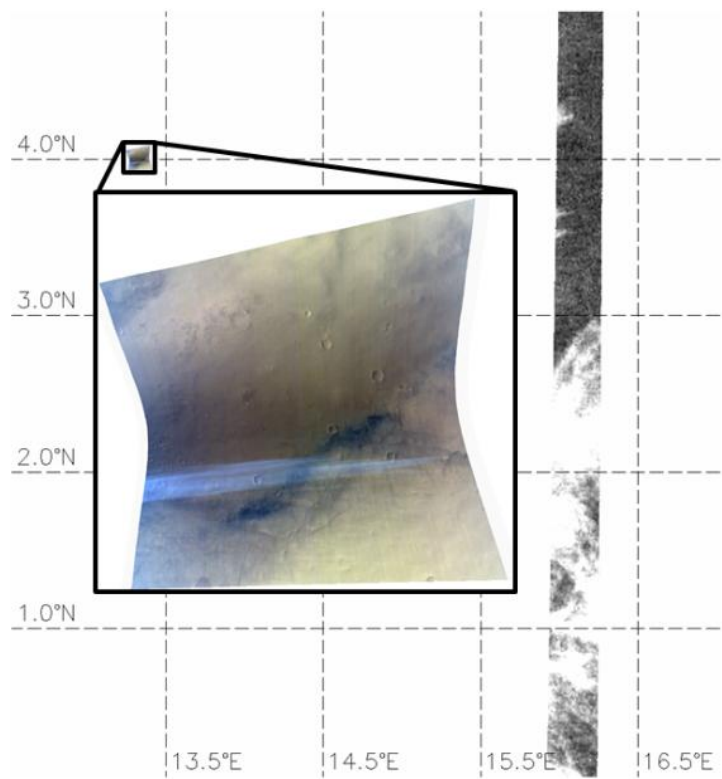

*Figure 15: Comparison of OMEGA (700 m per pixel) and CRISM (20 meters per pixel) observations of the same area containing $CO_2$ clouds at the same local time (15 h). Both observations have been obtained at Ls 31° in 2008. The OMEGA observation # 5285_3 covers the longitudinal band between 15.9°E and 16.3°E, while the small targeted CRISM observation # FRT9E67 is centered on 13.35°E at 4°N (see insert for a zoom of the CRISM image: CRISM only captures a small subset of large areas with $CO_2$ clouds). The OMEGA observation corresponds to the intensity at 4.24 µm (see Montmessin et al., 2007), whereas the CRISM image is a RGB composite at ~ 0.55 µm. The fine structure of the cloud is not resolved at OMEGA spatial resolution, and most of the cloud appears as "rounded isolated masses irregularly arranged resembling terrestrial alto- or cirro-cumulus" (citation from Montmessin et al., 2007). The fine structure revealed by CRISM show filaments in the east-west direction, a morphology resembling that of terrestrial cirrus, in agreement with HRSC observations.*

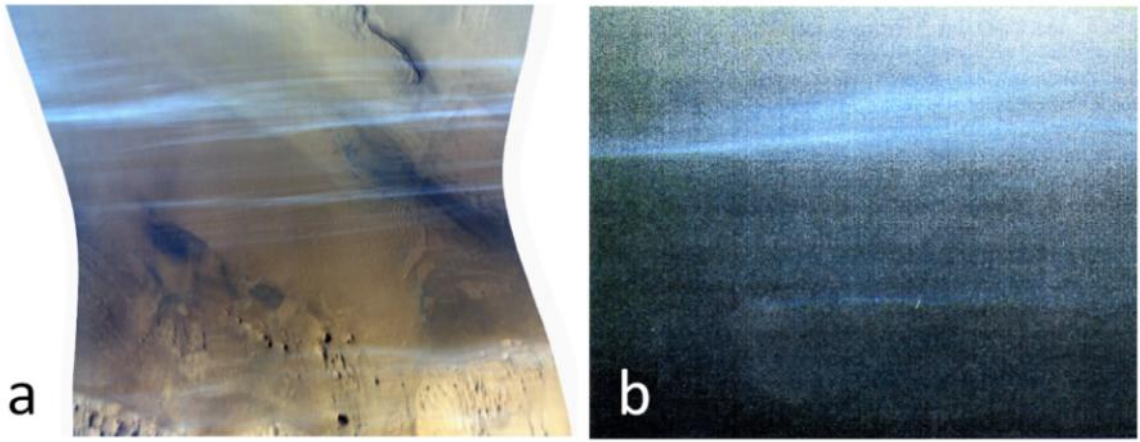

*Figure 16: Comparison between a CRISM observation of a $CO_2$ clouds (a) and a Pathfinder observation of a predawn cloud interpreted as a $CO_2$ cloud (b) (image from (Clancy & Sandor 1998)). Both clouds show a very similar morphology, independently of their viewing geometry (orbital observation for "a", observation from the surface for "b"). We can also notice that the cloud color is similar in both images, although RGB schemes have been independently computed.*

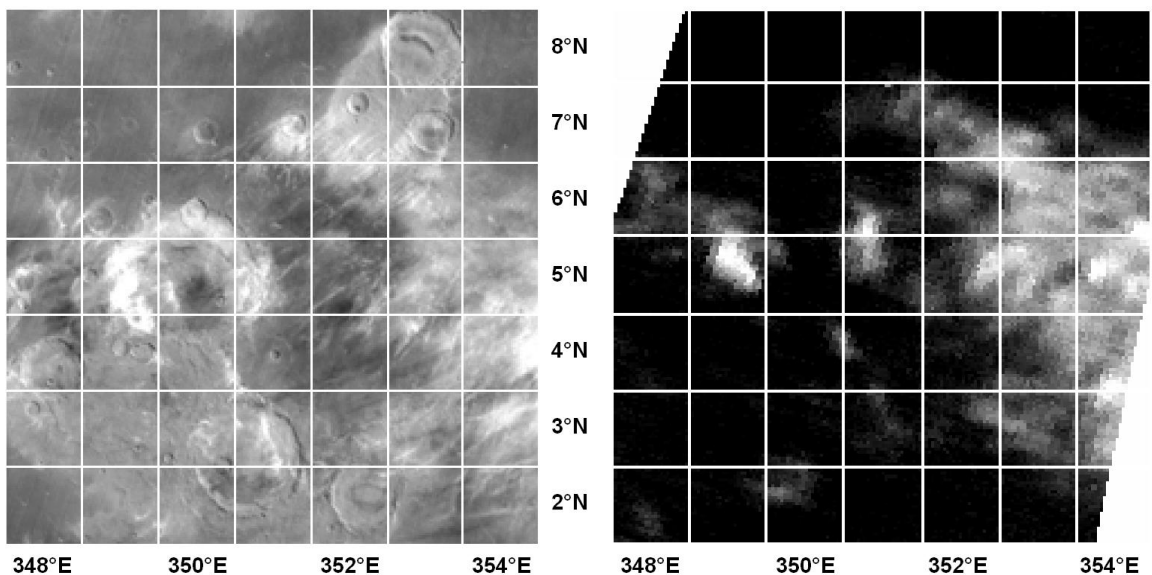

*Figure 17: HRSC (left) and OMEGA (right) observations of the same $CO_2$ cloud (orbite #7914). This $CO_2$ cloud has been observed in the morning at a local time of 9.5. The OMEGA observation has been obtained with a low spatial resolution of 6 km. In the OMEGA observation, the cloud appear similar to the "cumuliform" clouds defined by (Montmessin et al. 2007) and then by (Määttänen et al. 2010) who gives this denomination for morning clouds only. The morphology in the HRSC observation is similar to previously reported HRSC observations (Scholten et al. 2010), i.e. cirrus-type, which demonstrates that OMEGA morning "low resolution" $CO_2$ clouds observations cannot be used to estimate cloud morphology.*

Observed cloud morphologies provide clues regarding potential cloud formation and evolution mechanisms. Our analysis shows that equatorial $CO_2$ cloud observations acquired so far either show cirrus-type morphologies or do not have sufficient spatial resolution to estimate cloud morphology. This conclusion is valid for morning and afternoon clouds during northern spring and summer. Calculations of the convective available potential energy performed by (Määttänen et al. 2010) show that the energy available in a 3x3 km $CO_2$ cloud will create a convective cell smaller than 10x10m to 100x100 m. The probability to observe such a convective cell with CRISM is low, first because it represents 1/1000 to 1/100 000 of the surface of the cloud, and secondly because it is at the limit of CRISM high spatial resolution (20 m). As we do not observe such a convective cell, our observations are consistent with this calculation implying a minor contribution of convection to these clouds. Martian equatorial $CO_2$ clouds are composed of filaments with a preferential East-West orientation. This orientation is similar to wind orientation at these altitudes (Angelats i Coll et al. 2004; Scholten et al. 2010). (Angelats i Coll et al. 2004; Scholten et al. 2010). $CO_2$ clouds could be initiated locally, e.g. by gravity waves (Spiga et al. 2010), and then stretched by the easterly mesospheric winds.

## 4. Summary and conclusions

In this paper, we developed a new method to detect $CO_2$ clouds with near-IR orbital data acquired by CRISM. The method is based on radiative transfer modeling results showing that $CO_2$ clouds have distinct spectral properties compared to other potential aerosols. These CRISM observations were compared to previously published OMEGA $CO_2$ cloud detections (Montmessin et al. 2007; Määttänen et al. 2010) and to new OMEGA observations acquired in 2009 and 2010. An extensive comparison with mesospheric ice clouds (altitudes greater than 60 km) observed in various dataset (TES/MOC (Clancy et al. 2007), HRSC (Scholten et al. 2010), SPICAM (Montmessin et al. 2006) and THEMIS (McConnochie et al. 2010)) was also implemented.

The new CRISM measurements reveal 54 $CO_2$ clouds located between 13°S and 9°N: 24 near Valles Marineris (250°E-310°E), 29 near Terra Meridiani (350°E-30°E), and one at 155°E. These clouds occur within areas and periods of previously known $CO_2$ cloud activity from OMEGA data, except for the 155°E cloud. CRISM $CO_2$ clouds are observed between $L_S$ 0° and 140°, with a maximum activity at $L_S$ 20°. Their effective radius typically ranges from about 0.5 to 2 µm and their optical depths are up to 0.3 for the thickest clouds.

At low to mid latitudes, the distribution of spectroscopic identifications of $CO_2$ clouds is at first order consistent with the distribution of high altitude (> 60 km) ice clouds, which shows that $CO_2$ clouds dominate the mesosphere of Mars. Differences are observed at second order; they are notably due to infrequent mesospheric water ice clouds, as revealed by a water ice cloud layer extending up to 80 km observed by OMEGA.

$CO_2$ clouds in CRISM afternoon observations are composed of filaments and resemble terrestrial cirrus clouds; filaments are frequently East-West oriented. This extends the result of (Scholten et al. 2010) based on HRSC afternoon observations to the full $L_S$ range of equatorial $CO_2$ cloud activity. A combined OMEGA/HRSC morning observation suggests that OMEGA morning observations have been previously misinterpreted as evidence for the existence of cumuliform – and hence potentially convective – $CO_2$ clouds.

The nature and origin of cloud condensation nuclei has not been addressed in this study. With the help of radiative transfer models of radially inhomogeneous particles, near-IR spectroscopy may hold information about the composition of condensation nuclei. In addition to nanophase ferric oxides coming from the surface and micrometeorites coming from space, water ice grains – present at $CO_2$ clouds altitudes according to our limb observation – could provide efficient cores on which to condense. Interestingly, spectral properties derived above $CO_2$ clouds contain a shallow feature at 3 µm resembling water-related features. Further modeling is however required to understand the origin (surface or cloud) of this spectral feature.

## 5. Acknowledgment

The authors would like to thank R. T. Clancy, Y. Langevin, F. Montmessin, J. Mustard, B. Schmitt and A. Spiga for helpful discussions during the preparation of this paper.